\documentclass{ws-ijmpa}
\usepackage[]{cite}
\begin{document}

\markboth{Zekovi\'{c} et al.} {Relativistic non-thermal
bremsstrahlung radiation}

%
\catchline{}{}{}{}{}
%

\title{RELATIVISTIC NON-THERMAL
BREMSSTRAHLUNG RADIATION
}

\author{VLADIMIR ZEKOVI\'{C},$^*$ BOJAN ARBUTINA, ALEKSANDRA DOBARD\v{Z}I\'{C}\\ and MARKO PAVLOVI\'{C} }

\address{Department of Astronomy, Faculty of Mathematics, University of Belgrade, Studentski trg 16,\\
11000 Belgrade, Serbia\\
\footnote{Corresponding author} vlada@matf.bg.ac.rs}

\maketitle

\begin{history}
\received{16 July 2013} \accepted{14 October 2013}
\end{history}

\begin{abstract}
By applying a method of virtual quanta we derive formulae for
relativistic non-thermal bremsstrahlung radiation from
relativistic electrons as well as from protons and heavier
particles with power-law momentum distribution $N(p)dp = k
p^{-q}dp.$ We show that emission which originates from an electron
scattering on an ion, represents the most significant component of
relativistic non-thermal bremsstrahlung. Radiation from an ion
scattering on electron, known as inverse bremsstrahlung, is shown
to be negligible in overall non-thermal bremsstrahlung emission.
These results arise from theory refinement, where we introduce the
dependence of relativistic kinetic energy of an incident particle,
upon the energy of scattered photon. In part, it is also a
consequence of a different mass of particles and relativistic
effects.

\end{abstract}

\ccode{PACS numbers:  03.50.-z, 41.60.-m, 78.70.Ck, 96.50.S-,
13.85.Tp, 97.80.Jp, 98.70.Qy}

\section{Introduction}

{\it Bremsstrahlung} or deceleration radiation (form German {\it bremsen} "to brake" and {\it Strahlung} "radiation") is important radiation mechanism in
laboratory and astrophysical plasmas. Since protons are generally less motile than electrons, bremsstrahlung or free-free emission is often considered as a radiation (photon emission) of decelerating electron in the Coulomb field of an ion. Even synchrotron radiation of an ultra-relativistic electron gyrating about field lines in the magnetic field is sometimes referred as magneto-bremsstrahlung.

Thermal bremsstrahlung is often observed in astrophysical sources such as H\textsc{ii} or ionized hydrogen regions (e.g. Orion nebula) in radio domain or in clusters of galaxies i.e. hot intercluster medium, in the X-rays. It is produced by electrons with thermal (Maxwell-Boltzmann) distribution.

Accelerated electrons often have power law distribution and as a result produce non-thermal bremsstrahlung radiation. Power law distribution of particles can be produced in shock waves, so non-thermal bremsstrahlung radiation could be observed in objects such as supernova remnants \cite{r12}. Also hard X-ray emission from clusters of galaxies in which accretion or merger shocks are present could be produced in part by non-thermal bremsstrahlung radiation \cite{r13,r14}.

In paper \cite{r15}, the role of inverse bremsstrahlung radiation is considered in shocked astrophysical plasmas, defined to be the emission of a single photon when a high-speed ion collides with an electron that is effectively at rest. Their conclusion was that inverse bremsstrahlung can be neglected in most models of shock acceleration in supernova remnants and similar sources. However, problem of inverse bremsstrahlung detection on scales $\geq$ 100 pc, distant from these discrete sources, remains open.

Although electron-electron bremsstrahlung is normally ignored in comparison to normal electron-ion bremsstrahlung, it was pointed out
that it can contribute to the hard X-ray emission from solar flares \cite{r16}. These authors recognized the growing importance of electron-electron bremsstrahlung at electron (and photon) energies above 300 keV.

In this paper we consider the relativistic non-thermal
bremsstrahlung. In the following section, by applying a method of
virtual quanta \cite{r1,r2} (see also Ref.~\citen{r3}), we shall
derive formulae for relativistic non-thermal bremsstrahlung
radiation from ultra-relativistic electrons as well as from
protons and heavier particles with power-law momentum distribution
$N(p)dp = k p^{-q}dp$.

\section{Analysis and Results}

\subsection{Relativistic Electron Bremsstrahlung}

Let us first consider bremsstrahlung radiation of a
relativistic electron. In the rest frame of the electron it
appears that a proton, or in general an ion with charge $Ze$,
moves rapidly towards the electron. Electrostatic field of this
ion is transformed into a transverse pulse, which to electron
appears as a pulse of electromagnetic radiation -- a {\it virtual
quanta}. The quanta or photon scatters of electron and produces
detectable bremsstrahlung radiation. In the primed (electron's)
frame, the spectrum of the pulse of virtual quanta has the form
(see Ref.~\citen{r4})
\begin{equation}
\frac{\mathrm{d}W'}{\mathrm{d}S'\mathrm{d}\omega '} = \frac{Z^2 e^2 c}{\pi ^2 b'^2 v^2}\Big(
\frac{b'\omega '}{\gamma v} \Big) ^2 K_1^2 \Big( \frac{b'\omega
'}{\gamma v} \Big) ,
\end{equation}
where $W$ is energy, $S$ surface element, $\omega$ circular
frequency, $b$ impact parameter, $c$ speed of light, $\gamma =
1/\sqrt{1-\beta ^2}$ the Lorentz factor, $\beta =v/c$, $v$ is velocity and $K_1(x)$ the modified Bessel function of order one.

Assuming elastic scattering in the low-energy limit ($\hbar \omega '
\ll m_e c^2$) we have
\begin{equation}
\frac{\mathrm{d}W'}{\mathrm{d}\omega '} = \sigma _\mathrm{T} \frac{\mathrm{d}W'}{\mathrm{d}S'\mathrm{d}\omega '} ,
\end{equation}
where $\sigma _\mathrm{T}$ is Thomson cross section. Since energy and
frequency transform identically under Lorentz transformations
$\mathrm{d}W/\mathrm{d}\omega =\mathrm{d}W'/\mathrm{d}\omega '$, transverse lengths are unchanged
$b=b'$, and $\omega = \gamma \omega ' (1+\beta \cos \theta ')
\approx \gamma \omega '$ on average, if the scattering is
forward-backward symmetric, in the laboratory frame we have
\begin{equation}
\frac{\mathrm{d}W}{\mathrm{d}\omega } = \frac{8Z^2e^6}{3\pi b^2 m_e^2 c^3 v^2} \Big(
\frac{b\omega }{\gamma ^2 v} \Big) ^2 K_1^2 \Big( \frac{b \omega
}{\gamma ^2 v} \Big).
\end{equation}

Emitted power per unit frequency of the single electron is
\begin{equation}
\frac{\mathrm{d}W}{\mathrm{d}t \mathrm{d}\omega } = 2 \pi c n_i \int _{b_\mathrm{min}}^\infty
\frac{\mathrm{d}W}{\mathrm{d}\omega } b \mathrm{d}b,
\end{equation}
where $n_i$ is ion number density and we shall set $b_\mathrm{min} =
\frac{ \hbar}{m_e v}$ \cite{r4,r5}. For a power-law  distribution
of electrons $N(E)\mathrm{d}E = K_e E^{-q}\mathrm{d}E$, where $q$ is energy index:
\begin{equation}
\frac{\mathrm{d}W}{\mathrm{d}t \mathrm{d}\omega \mathrm{d}V} = \int _0 ^\infty \frac{\mathrm{d}W}{\mathrm{d}t \mathrm{d}\omega }
N(E)\mathrm{d}E
\end{equation}
where, as usual, the integration limits are set from 0 to $\infty$ in approximation of ultra-relativistic particles. We derive our solution in a more realistic way, by taking into account cosmic rays of lower energies and by not making the assumption that the particle's speed is always close to $c$, which leads to a more complex derivation. We use momentum instead of the energy distribution in (5) and we also introduce finite integration limits:
\begin{eqnarray}
\frac{\mathrm{d}W}{\mathrm{d}t \mathrm{d}\omega \mathrm{d}V} &=& \int _{p_{\mathrm{min}}} ^{p_{\mathrm{max}}} \frac{\mathrm{d}W}{\mathrm{d}t \mathrm{d}\omega }
N(p)\mathrm{d} p, \nonumber \\
\frac{\mathrm{d}W}{\mathrm{d}t \mathrm{d}\omega \mathrm{d}V} &=& \frac{16Z^2e^6}{3 m_e^2 c^3} n_i k_e
\int _{p_{\mathrm{min}}} ^{p_{\mathrm{max}}} \frac{1}{v} p^{-q} \mathrm{d}p \int _x ^\infty y K_1^2(y)
\mathrm{d}y,
\end{eqnarray}
where $k_e$ is constant of the momentum distribution of electrons $N(p) \mathrm{d}p = k_e p^{-q} \mathrm{d}p$ and $k_e = K_e c^{1-q}$; the momentum boundaries $p_{\mathrm{min}}$ and $p_{\mathrm{max}}$ will be discussed in the following text; collision parameter is $b_{\mathrm{min}} = \frac{ \hbar} {m_e v}$; $y=\frac{\omega b}{\gamma^2  v}$ and $x=\frac{\omega b_{\mathrm{min}}} {\gamma^2 v}= \frac{m_e \hbar \omega }{p^2}$. If we express $(pc)$ in terms of $x$
\begin{eqnarray}
pc &=& \sqrt{\frac{m_e c^2 \hbar \omega}{x}},\nonumber \\
(pc)\mathrm{d}(pc) &=& -\frac{1}{2} \frac{m_e c^2 \hbar \omega}{x^2} \mathrm{d}x \nonumber
\end{eqnarray}
and change it in (6), we derive
\begin{eqnarray}
\frac{\mathrm{d}W}{\mathrm{d}t \mathrm{d}\omega \mathrm{d}V} &=& \frac{16Z^2e^6}{3 m_e^2 c^4} n_i (k_e c^{q-1}) (m_e c^2 \hbar \omega)^{(1-q)/2} \cdot \int _{x_{\mathrm{min}}} ^{x_{\mathrm{max}}} \frac{1}{2} x^{(q-3)/2} \sqrt{1 + \frac{m_e c^2}{\hbar \omega} x}\nonumber \\
&\cdot & \Big[xK_0(x)K_1(x)-\frac{1}{2}x^2(K_1^2(x)-K_0^2(x))\Big] \mathrm{d}x,
\end{eqnarray}
with integration limits $x_{\mathrm{min}} = \frac{m_e \hbar \omega }{p_{\mathrm{max}}^2}$ and $x_{\mathrm{max}} = \frac{m_e \hbar \omega }{p_{\mathrm{min}}^2}.$
In the above derivation we used identity
\begin{equation}
\int _x ^\infty y K_1^2(y) \mathrm{d}y =
xK_0(x)K_1(x)-\frac{1}{2}x^2(K_1^2(x)-K_0^2(x)).
\end{equation}

We now derive analytical solution to (7), but first let us discuss
the integration boundaries. Varying the momentum upper limit
$p_{\mathrm{max}}$ from $\sim 10^{15} \rm eV/c$ to $\infty$ will
always give values of $x_{\mathrm{min}}$ very close to zero, so we
use $p_{\mathrm{max}} = \infty$ and $x_{\mathrm{min}} = 0.$ The
lower limit of momentum $p_{\mathrm{min}},$ however small it is,
will always give $x_{\mathrm{max}} \ll 1$ for photon energies in
the scope of Thomson scattering. This enables us to use
approximation of relation (8) (see Ref.9)
\begin{equation}
xK_0(x)K_1(x)-\frac{1}{2}x^2(K_1^2(x)-K_0^2(x)) \approx \ln \left( \frac{0.684}{x} \right),
\end{equation}
where ${\rm \bf e}^{-\sqrt{\curlyvee}/2}\approx 0.684$ and
$\curlyvee \approx 0.577$ is Euler's gamma constant, which then
results in analytically more solvable expression than (7)
\begin{eqnarray}
\frac{\mathrm{d}W}{\mathrm{d}t \mathrm{d}\omega \mathrm{d}V} &=& \frac{16Z^2e^6}{3 m_e^2 c^4} n_i K_e (m_e c^2 \hbar \omega)^{(1-q)/2} \nonumber \\
&\cdot & \int _0 ^{x_{\mathrm{max}}} \frac{1}{2} x^{(q-3)/2} \sqrt{1 + \frac{m_e c^2}{\hbar \omega} x} \cdot \ln \left( \frac{0.684}{x} \right) \mathrm{d}x.
\end{eqnarray}

Here we introduce the new method of solving complicated integrals in a way of getting very precise solutions analytically, only if limiting conditions can be defined so that integrals can take much simpler form and become solvable. As a first step, we make use of the integral mean value theorem which states that
\begin{equation}
\overline{f(x)} = \frac{1}{b-a} \cdot \int _a ^b f(x) \mathrm{d}x
\end{equation}
and we define some mean value $\overline{x},$ so that $f(\overline{x}) = \overline{f(x)}.$ Applying the theorem to relation (10) leads to
\begin{eqnarray}
\frac{\mathrm{d}W}{\mathrm{d}t \mathrm{d}\omega \mathrm{d}V} &=& \frac{16Z^2e^6}{3 m_e^2 c^4} n_i K_e (m_e c^2 \hbar \omega)^{(1-q)/2} \cdot \frac{1}{2} \cdot x_{\mathrm{max}} \cdot f(\overline{x}) \\
f(\overline{x}) &=& \overline{x}^{(q-3)/2} \sqrt{1 + \frac{m_e c^2}{\hbar \omega} \overline{x}} \cdot \ln \left( \frac{0.684}{\overline{x}} \right). \nonumber
\end{eqnarray}
As a second step, we want to make relation between $\overline{x}$ and $x_{\mathrm{max}},$ so we define the two boundary cases of (10). First, when $\frac{m_e c^2}{\hbar \omega} x \gg 1,$ so that $\sqrt{1 + \frac{m_e c^2}{\hbar \omega} x} \approx \sqrt{\frac{m_e c^2}{\hbar \omega} x}.$ This is valid for low energy photons and ideally when $\hbar \omega \rightarrow 0.$ In this limit relation (10) reduces to
\begin{eqnarray}
\frac{\mathrm{d}W}{\mathrm{d}t \mathrm{d}\omega \mathrm{d}V} \bigg| _{\hbar \omega \rightarrow 0} &=& \frac{16Z^2e^6}{3 m_e^2 c^4} n_i K_e (m_e c^2)^{1-q/2} (\hbar \omega)^{-q/2} \\
&\cdot & \int _0 ^{x_{\mathrm{max}}} \frac{1}{2} x^{q/2-1} \ln \left( \frac{0.684}{x} \right) \mathrm{d}x. \nonumber
\end{eqnarray}
Second case is defined over domain of high energy photons ($\hbar \omega \rightarrow \infty$), when $\frac{m_e c^2}{\hbar \omega} x \ll 1$ and $\sqrt{1 + \frac{m_e c^2}{\hbar \omega} x} \approx 1,$ which leads to the high energy limit
\begin{eqnarray}
\frac{\mathrm{d}W}{\mathrm{d}t \mathrm{d}\omega \mathrm{d}V} \bigg| _{\hbar \omega \rightarrow \infty} &=& \frac{16Z^2e^6}{3 m_e^2 c^4} n_i K_e (m_e c^2 \hbar \omega)^{(1-q)/2} \\
&\cdot & \int _0 ^{x_{\mathrm{max}}} \frac{1}{2} x^{(q-3)/2} \ln \left( \frac{0.684}{x} \right) \mathrm{d}x. \nonumber
\end{eqnarray}
Solutions to integrals (13) and (14) are now straightforward and are given by (15) and (16) respectively:
\begin{eqnarray}
\frac{\mathrm{d}W}{\mathrm{d}t \mathrm{d}\omega \mathrm{d}V} \bigg| _{\hbar \omega \rightarrow 0} &=& \frac{16Z^2e^6}{3 m_e^2 c^4} n_i K_e (m_e c^2)^{1-q/2} (\hbar \omega)^{-q/2} \cdot x_{\mathrm{max}}^{q/2} \\
&\cdot & \frac{1}{q} \left[ \ln \left( \frac{0.684}{x_{\mathrm{max}}} \right) + \frac{2}{q} \right], \nonumber \\
\frac{\mathrm{d}W}{\mathrm{d}t \mathrm{d}\omega \mathrm{d}V} \bigg| _{\hbar \omega \rightarrow \infty} &=& \frac{16Z^2e^6}{3 m_e^2 c^4} n_i K_e (m_e c^2)^{(1-q)/2} (\hbar \omega)^{(1-q)/2} \cdot x_{\mathrm{max}}^{(q-1)/2} \\
&\cdot & \frac{1}{q-1} \left[ \ln \left( \frac{0.684}{x_{\mathrm{max}}} \right) + \frac{2}{q-1} \right]. \nonumber
\end{eqnarray}
We now make use of the first boundary solution (15) and equate it
with relation (12) in the limit of very low photon energies
\begin{equation}
\overline{x}^{q/2-1} \ln \frac{0.684}{\overline{x}} \equiv x_{\mathrm{max}}^{q/2-1} \frac{2}{q} \left[ \ln \left( \frac{0.684}{x_{\mathrm{max}}} \right) + \frac{2}{q} \right] = x_{\mathrm{max}}^{q/2-1} \frac{2}{q} \ln \left( \frac{0.684}{x_{\mathrm{max}} \cdot {\rm \bf e}^{-2/q}} \right).
\end{equation}
Because in most cases spectral index lies in the range $2\leq q\leq 3,$ we can expand the series
\begin{equation}
\left( {\rm \bf e}^{-2/q} \right) ^{q/2-1} = {\rm \bf e}^{2/q-1} \approx \frac{2}{q} \nonumber
\end{equation}
and substitute it in (17) to get
\begin{equation}
\overline{x}^{q/2-1} \cdot \ln \frac{0.684}{\overline{x}} = \left( x_{\mathrm{max}} \cdot {\rm \bf e}^{-2/q} \right) ^{q/2-1} \cdot \ln \left( \frac{0.684}{x_{\mathrm{max}} \cdot {\rm \bf e}^{-2/q}} \right), \nonumber
\end{equation}
which then results in relation for low energy photons
\begin{equation}
\overline{x} = x_{\mathrm{max}} \cdot {\rm \bf e}^{-2/q}.
\end{equation}
We now equate the second boundary solution (16) with relation (12)
written in the limit of very high photon energies
\begin{equation}
\overline{x}^{(q-3)/2} \cdot \ln \frac{0.684}{\overline{x}} \equiv x_{\mathrm{max}}^{(q-3)/2} \cdot \frac{2}{q-1} \ln \left( \frac{0.684}{x_{\mathrm{max}} \cdot {\rm \bf e}^{-2/(q-1)}} \right).
\end{equation}
As previously, we can represent $2/(q-1)$ with a series expansion
\begin{equation}
\left( {\rm \bf e}^{-2/(q-1)} \right) ^{(q-3)/2} = {\rm \bf e}^{2/(q-1)-1} \approx \frac{2}{q-1} \nonumber
\end{equation}
and change it in (19) to get relation which holds in the range of high energy photons
\begin{equation}
\overline{x} = x_{\mathrm{max}} \cdot {\rm \bf e}^{-2/(q-1)}.
\end{equation}
As a balance between low and high energy solutions, we make use of the geometric mean of (18) and (20) to get our analytical solution to fit the whole range of photon energies
\begin{equation}
\overline{x} = \sqrt{x_{\mathrm{max}}^2 \cdot {\rm \bf e}^{-2/q} \cdot {\rm \bf e}^{-2/(q-1)}} = x_{\mathrm{max}} \cdot {\rm \bf e}^{(1-2q)/(q(q-1))}.
\end{equation}
We use this relation and after substituting it in (12), we derive
\begin{eqnarray}
\frac{\mathrm{d}W}{\mathrm{d}t \mathrm{d}\omega \mathrm{d}V} &=& \frac{16Z^2e^6}{3 m_e^2 c^4} n_i K_e (m_e c^2 \hbar \omega)^{(1-q)/2} \cdot \frac{1}{2} x_{\mathrm{max}}^{(q-1)/2} \cdot {\rm \bf e}^{Q(q-3)/2} \\
&\cdot & \sqrt{1 + \frac{m_e c^2}{\hbar \omega} x_{\mathrm{max}} \cdot {\rm \bf e}^{Q}} \cdot \left[ \ln \left( \frac{0.684}{x_{\mathrm{max}}} \right) + \frac{1}{q} + \frac{1}{q-1} \right] , \nonumber
\end{eqnarray}
where $Q = \frac{1-2q}{q(q-1)} = - \left[ \frac{1}{q} + \frac{1}{q-1} \right].$

If we now express the lowest value of momentum $p_{\mathrm{min}}$, that we take into account in $x_{\mathrm{max}},$ in terms of electron's kinetic energy $T_e$
\begin{equation}
x_{\mathrm{max}} = \hbar \omega \frac{m_e}{p_{\mathrm{min}}^2} = \hbar \omega \cdot \frac{m_e c^2}{T_e^2 + 2 m_e c^2 T_e}
\end{equation}
and if we define that kinetic energy of a particle is related to the energy of a photon emitted by that particle as
\begin{equation}
T_e = \zeta_e \hbar \omega,
\end{equation}
so that
\begin{equation}
x_{\mathrm{max}} = \hbar \omega \cdot \frac{m_e c^2}{( \zeta_e \hbar \omega )^2 + 2 \zeta_e \hbar \omega m_e c^2} = \frac{1}{2 \zeta_e} \cdot \frac{1}{1 + \frac{1}{2} \frac{\zeta_e \hbar \omega}{m_e c^2}},
\end{equation}
where $\zeta_e$ is generally not constant and can be a function of $\hbar \omega,$ then we can rewrite expression (22) completely as a function of photon energy
\begin{eqnarray}
\frac{\mathrm{d}W}{\mathrm{d}t \mathrm{d}\omega \mathrm{d}V} &=& \frac{16Z^2e^6}{3 m_e^2 c^4} n_i K_e (\zeta_e \hbar \omega)^{1-q} \cdot \left( 1 + 2 \frac{m_e c^2}{\zeta_e \hbar \omega} \right) ^{1-q/2} \frac{1}{2} \cdot {\rm \bf e}^{Q(q-3)/2} \\
&\cdot & \sqrt{1 + \frac{{\rm \bf e}^{Q} - 2 \left( 2 + \frac{\zeta_e \hbar \omega}{m_e c^2} \right) }{\left( 2 + \frac{\zeta_e \hbar \omega}{m_e c^2} \right) ^2}} \cdot \left[ \ln \left( 0.684 \cdot \zeta_e \left( 2 + \frac{\zeta_e \hbar \omega}{m_e c^2} \right) \right) + \frac{1}{q} + \frac{1}{q-1} \right] . \nonumber
\end{eqnarray}
This relation represents an approximate analytical solution to electron non-thermal bremsstrahlung emission, which holds for all photon energies in the scope of Thomson scattering.

There is a break in the spectrum of electron non-thermal bremsstrahlung emission around $\frac{\zeta_e \hbar \omega}{m_e c^2} \approx 2$. As an approximation of (26), the two power laws are then given for low and high energy parts of the spectrum by (27) and (28) respectively
\begin{eqnarray}
\frac{\mathrm{d}W}{\mathrm{d}t \mathrm{d}\omega \mathrm{d}V} &=& \frac{16Z^2e^6}{3 m_e^2 c^4} n_i K_e (2 m_e c^2)^{1-q/2}(\zeta_e \hbar \omega)^{-q/2} \\
&\cdot & \frac{1}{4} {\rm \bf e}^{Q(q-2)/2} \left[ \ln \left( 0.684 \cdot 2 \zeta_e \right) + \frac{1}{q} + \frac{1}{q-1} \right] , \nonumber \\
\frac{\mathrm{d}W}{\mathrm{d}t \mathrm{d}\omega \mathrm{d}V} &=& \frac{16Z^2e^6}{3 m_e^2 c^4} n_i K_e (\zeta_e \hbar \omega)^{1-q} \\
&\cdot & \frac{1}{2} {\rm \bf e}^{Q(q-3)/2} \left[ \ln \left( 0.684 \cdot 2 \zeta_e \right) + \ln \left( 1 + \frac{1}{2} \frac{\zeta_e \hbar \omega}{m_e c^2} \right) + \frac{1}{q} + \frac{1}{q-1} \right] \nonumber
\end{eqnarray}
and the break point is found as
\begin{equation}
\hbar \omega _{\rm b} \approx \frac{2 m_e c^2}{\zeta_e }.
\end{equation}

\subsection{Relativistic Ion Bremsstrahlung}

In the case of relativistic proton or heavier ion with charge $Ze$, mass
$Am_p$\footnote{$A$ is the mass number; for hydrogen i.e. proton
it is 1, for helium i.e. $\alpha$-particle it is 4 (assuming
nearly equal mass of proton and neutron), etc.} and energy $E$ we are dealing with the scattering of quanta or photon over electron in the laboratory frame, assuming that during the interaction the electron thermal velocity is small enough, so that it is
practically at rest and that the photon energy is small compared to electron's rest energy, i.e.
$kT_e \ll \hbar \omega \ll m_e c^2 \ll E$. If this is not the case the scattering would
not be Thomson's but Compton's and Inverse Compton's.

Since now we do not have to transform from primed to laboratory
frame, we have
\begin{equation}
\frac{\mathrm{d}W}{\mathrm{d}\omega } = \frac{8Z^2e^6}{3\pi b^2 m_e^2 c^3 v^2} \Big(
\frac{b\omega }{\gamma v} \Big) ^2 K_1^2 \Big( \frac{b \omega
}{\gamma  v} \Big).
\end{equation}
If again $b_{\mathrm{min}} = \frac{ \hbar}{m_e v}$, $y=\frac{\omega
b}{\gamma  v}$ and $x=\frac{\omega b_{\mathrm{min}}} {\gamma v}= \hbar \omega \frac{A m_p}{m_e} \frac{ E }{(pc)^2}$ ($E=\gamma A m_p c^2=\sqrt{(pc)^2+(A m_p c^2)^2}$) we have
\begin{equation}
\frac{\mathrm{d}W}{\mathrm{d}t \mathrm{d}\omega \mathrm{d}V} = \frac{16Z^2e^6}{3 m_e^2 c^3} n_e k_i
\int _{p_{\mathrm{min}}} ^{p_{\mathrm{max}}} \frac{1}{v} p^{-q} \mathrm{d}p \int _x ^\infty y K_1^2(y)
\mathrm{d}y,
\end{equation}
where $k_i$ is now constant of the momentum distribution of ions $N(p) \mathrm{d}p = k_i p^{-q} \mathrm{d}p$ and $k_i = K_i c^{1-q}.$ If we now express $(pc)^2$ in terms of $x$
\begin{eqnarray}&(pc)^4& \left( \frac{m_e}{A m_p \hbar \omega} \right)^2 x^2 - (pc)^2 - (A m_p c^2)^2 = 0,\nonumber \\
&(pc)^2& = \frac{1+\sqrt{1 + \left( 2 \frac{m_e c^2}{\hbar \omega} x \right)^2}}{2 \left( \frac{m_e}{A m_p \hbar \omega} x \right)^2}, \nonumber \\
&\mathrm{d}(pc)^2& = \left[ -\frac{1 + \sqrt{1 + \left( 2 \frac{m_e c^2}{\hbar \omega} x \right)^2}}{\left( \frac{m_e}{A m_p \hbar \omega} \right)^2 x^3} + \frac{2 (A m_p c^2)^2}{x \sqrt{1 + \left( 2 \frac{m_e c^2}{\hbar \omega} x \right)^2}} \right] \mathrm{d}x = 2 c^2 p \mathrm{d}p\nonumber
\end{eqnarray}
and change it in (31), we get
\begin{eqnarray}
\frac{\mathrm{d}W}{\mathrm{d}t \mathrm{d}\omega \mathrm{d}V} &=& \frac{16Z^2e^6}{3 m_e^2 c^4} n_e (k_i c^{q-1}) \left(\frac{A m_p \hbar \omega}{m_e}\right)^{1-q} \int _{x_{\mathrm{min}}} ^{x_{\mathrm{max}}} \left( \frac{2 x^2}{1 + \sqrt{1 + \left( 2 \frac{m_e c^2}{\hbar \omega} x \right)^2}} \right)^{q/2-1}\nonumber \\
&\cdot & \left[ 1 - \frac{\frac{1}{2} \cdot \left( 2 \frac{m_e c^2}{\hbar \omega} x \right)^2}{\sqrt{1 + \left( 2 \frac{m_e c^2}{\hbar \omega} x \right)^2} \cdot \left( 1 + \sqrt{1 + \left( 2 \frac{m_e c^2}{\hbar \omega} x \right)^2} \right) } \right] \\
&\cdot & \Big[xK_0(x)K_1(x)-\frac{1}{2}x^2(K_1^2(x)-K_0^2(x))\Big] \mathrm{d}x.\nonumber
\end{eqnarray}

As in previous section, we use approximation (9) and discussion about the integration boundaries to derive a more analytically solvable expression
\begin{eqnarray}
\frac{\mathrm{d}W}{\mathrm{d}t \mathrm{d}\omega \mathrm{d}V} &=& \frac{16Z^2e^6}{3 m_e^2 c^4} n_e K_i \left(\frac{A m_p \hbar \omega}{m_e}\right)^{1-q} \int _0 ^{x_{\mathrm{max}}} \left( \frac{2 x^2}{1 + \sqrt{1 + \left( 2 \frac{m_e c^2}{\hbar \omega} x \right)^2}} \right)^{q/2-1}\nonumber \\
&\cdot & \left[ 1 - \frac{\frac{1}{2} \cdot \left( 2 \frac{m_e c^2}{\hbar \omega} x \right)^2}{\sqrt{1 + \left( 2 \frac{m_e c^2}{\hbar \omega} x \right)^2} \cdot \left( 1 + \sqrt{1 + \left( 2 \frac{m_e c^2}{\hbar \omega} x \right)^2} \right) } \right] \cdot \ln \frac{0.684}{x} \mathrm{d}x.
\end{eqnarray}
We again make use of the integral mean value theorem and as in the case of electrons we define some mean value $\overline{x},$ so that $f(\overline{x}) = \overline{f(x)}.$ Applying this to relation (33) leads to
\begin{eqnarray}
\frac{\mathrm{d}W}{\mathrm{d}t \mathrm{d}\omega \mathrm{d}V} &=& \frac{16Z^2e^6}{3 m_e^2 c^4} n_e K_i \left(\frac{A m_p \hbar \omega}{m_e}\right)^{1-q} \cdot x_{\mathrm{max}} \cdot f(\overline{x}) ,\\
f(\overline{x}) &=& \left(\frac{2 \overline{x}^2}{1 + \sqrt{1 + \left( 2 \frac{m_e c^2}{\hbar \omega} \overline{x} \right)^2}} \right)^{q/2-1} \nonumber \\
&\cdot & \left[ 1 - \frac{1}{2} \cdot \frac{\left( 2 \frac{m_e c^2}{\hbar \omega} \overline{x} \right)^2}{\sqrt{1 + \left( 2 \frac{m_e c^2}{\hbar \omega} \overline{x} \right)^2} \cdot \left( 1 + \sqrt{1 + \left( 2 \frac{m_e c^2}{\hbar \omega} \overline{x} \right)^2} \right) } \right] \cdot \ln \frac{0.684}{\overline{x}}. \nonumber
\end{eqnarray}
To make a connection to $x_{\mathrm{max}},$ we introduce the two boundary cases of (33), first when $\left( 2 \frac{m_e c^2}{\hbar \omega} x \right) ^2 \gg 1$ and (33) then simplifies to an approximate relation
\begin{eqnarray}
\frac{\mathrm{d}W}{\mathrm{d}t \mathrm{d}\omega \mathrm{d}V} &\approx & \frac{16Z^2e^6}{3 m_e^2 c^4} n_e K_i \left(\frac{A m_p \hbar \omega}{m_e}\right)^{1-q} \int _0 ^{x_{\mathrm{max}}} \left(\frac{2 x^2}{1 + 2 \frac{m_e c^2}{\hbar \omega} x} \right)^{q/2-1}\nonumber \\
&\cdot & \left[ 1 - \frac{1}{2} \cdot \frac{2 \frac{m_e c^2}{\hbar \omega} x}{1 + 2 \frac{m_e c^2}{\hbar \omega} x} \right] \cdot \ln \frac{0.684}{x} \ \mathrm{d}x. \nonumber
\end{eqnarray}
We can make the further simplification, which is the subcase when $2 \frac{m_e c^2}{\hbar \omega} x \gg 1$ or the low energy limit when $\hbar \omega \rightarrow 0$:
\begin{eqnarray}
\frac{\mathrm{d}W}{\mathrm{d}t \mathrm{d}\omega \mathrm{d}V} \bigg| _{\hbar \omega \rightarrow 0} &=& \frac{16Z^2e^6}{3 m_e^2 c^4} n_e K_i (A m_p c^2)^{1-q/2} \left(\frac{A m_p \hbar \omega}{m_e}\right)^{-q/2} \nonumber \\
&\cdot & \int _0 ^{x_{\mathrm{max}}} \frac{1}{2} \ x^{q/2-1} \ln \frac{0.684}{x} \ \mathrm{d}x.
\end{eqnarray}
In the second case $(2 \frac{m_e c^2}{\hbar \omega} x)^2 \ll 1$ when $\hbar \omega \rightarrow \infty$, (33) becomes
\begin{equation}
\frac{\mathrm{d}W}{\mathrm{d}t \mathrm{d}\omega \mathrm{d}V} \bigg| _{\hbar \omega \rightarrow \infty} = \frac{16Z^2e^6}{3 m_e^2 c^4} n_e K_i \left(\frac{A m_p \hbar \omega}{m_e}\right)^{1-q} \cdot \int _0 ^{x_{\mathrm{max}}} x^{q-2} \ln \frac{0.684}{x} \ \mathrm{d}x.
\end{equation}
As in the case of relativistic electron bremsstrahlung, solutions to (35) and (36) are straightforward and are given by (37) and (38) respectively:
\begin{eqnarray}
\frac{\mathrm{d}W}{\mathrm{d}t \mathrm{d}\omega \mathrm{d}V}\bigg| _{\hbar \omega \rightarrow 0} &=& \frac{16Z^2e^6}{3 m_e^2 c^4} n_e K_i (A m_p c^2)^{1-q/2} \left(\frac{A m_p \hbar \omega}{m_e}\right)^{-q/2} \cdot x_{\mathrm{max}}^{q/2} \\
&\cdot & \frac{1}{q} \left[ \ln \left( \frac{0.684}{x_{\mathrm{max}}} \right) + \frac{2}{q} \right], \nonumber \\
\frac{\mathrm{d}W}{\mathrm{d}t \mathrm{d}\omega \mathrm{d}V}\bigg| _{\hbar \omega \rightarrow \infty} &=& \frac{16Z^2e^6}{3 m_e^2 c^4} n_e K_i \left(\frac{A m_p \hbar \omega}{m_e}\right)^{1-q} \cdot x_{\mathrm{max}}^{q-1} \\
&\cdot & \frac{1}{q-1} \left[ \ln \left( \frac{0.684}{x_{\mathrm{max}}} \right) + \frac{1}{q-1} \right] . \nonumber
\end{eqnarray}
We first make use of boundary solution (37) and equate it with
relation (34) in the limit of very low photon energies when
\begin{equation}
1 + \sqrt{1 + \left( 2 \frac{m_e c^2}{\hbar \omega} \overline{x} \right)^2} \approx \sqrt{1 + \left( 2 \frac{m_e c^2}{\hbar \omega} \overline{x} \right)^2} \approx 2 \frac{m_e c^2}{\hbar \omega} \overline{x}, \nonumber
\end{equation}
to get relation between $\overline{x}$ and $x_{\mathrm{max}}$
\begin{equation}
\overline{x}^{q/2-1} \ln \frac{0.684}{\overline{x}} \equiv x_{\mathrm{max}}^{q/2-1} \frac{2}{q} \left[ \ln \left( \frac{0.684}{x_{\mathrm{max}}} \right) + \frac{2}{q} \right] = x_{\mathrm{max}}^{q/2-1} \frac{2}{q} \ln \left( \frac{0.684}{x_{\mathrm{max}} \cdot {\rm \bf e}^{-2/q}} \right).
\end{equation}
It is argued in previous section that we can expand series
\begin{equation}
\left( {\rm \bf e}^{-2/q} \right) ^{q/2-1} = {\rm \bf e}^{2/q-1} \approx \frac{2}{q} \nonumber
\end{equation}
and change it in (39) to derive
\begin{equation}
\overline{x}^{q/2-1} \cdot \ln \frac{0.684}{\overline{x}} = \left( x_{\mathrm{max}} \cdot {\rm \bf e}^{-2/q} \right) ^{q/2-1} \cdot \ln \left( \frac{0.684}{x_{\mathrm{max}} \cdot {\rm \bf e}^{-2/q}} \right), \nonumber
\end{equation}
which then results in a connection
\begin{equation}
\overline{x} = x_{\mathrm{max}} \cdot {\rm \bf e}^{-2/q}.
\end{equation}
We then equate (38) with relation (34) written in the limit of
very high photon energies when $1 + \sqrt{1 + \left( 2 \frac{m_e
c^2}{\hbar \omega} \overline{x} \right) ^2} \approx 2$ and we get
relation between $\overline{x}$ and $x_{\mathrm{max}}$
\begin{equation}
\overline{x}^{q-2} \ln \frac{0.684}{\overline{x}} \equiv \frac{x_{\mathrm{max}}^{q-2}}{q-1} \left[ \ln \left( \frac{0.684}{x_{\mathrm{max}}} \right) + \frac{1}{q-1} \right] = \frac{x_{\mathrm{max}}^{q-2}}{q-1} \ln \left( \frac{0.684}{x_{\mathrm{max}} {\rm \bf e}^{-1/(q-1)}} \right).
\end{equation}
As previously, we can represent $1/(q-1)$ as series expansion
\begin{equation}
\left( {\rm \bf e}^{-1/(q-1)} \right) ^{q-2} = {\rm \bf e}^{1/(q-1)-1} \approx \frac{1}{q-1} \nonumber
\end{equation}
and change it in (41) to get relation which holds for high energy photons
\begin{equation}
\overline{x} = x_{\mathrm{max}} \cdot {\rm \bf e}^{-1/(q-1)}.
\end{equation}
To get our analytical solution to fit the whole range of photon energies, we make compromise again by using the geometric mean of (40) and (42)
\begin{equation}
\overline{x} = \sqrt{x_{\mathrm{max}}^2 \cdot {\rm \bf e}^{-2/q} \cdot {\rm \bf e}^{-1/(q-1)}} = x_{\mathrm{max}} \cdot {\rm \bf e}^{(2-3q)/(2q(q-1))}.
\end{equation}

We now use this relation and after including it in (34), we derive an approximate analytical solution to relativistic ion (or inverse) non-thermal bremsstrahlung emission, which holds for all photon energies in the scope of Thomson scattering:
\begin{eqnarray}
\frac{\mathrm{d}W}{\mathrm{d}t \mathrm{d}\omega \mathrm{d}V} &=& \frac{16 Z^2 e^6}{3 m_e^2 c^4} n_e K_i \left(\frac{A m_p \hbar \omega}{m_e}\right)^{1-q} x_{\mathrm{max}}^{q-1} \left(\frac{2 \cdot {\rm \bf e}^{2Q}}{1 + \sqrt{1 + \left( 2 \frac{m_e c^2}{\hbar \omega} x_{\mathrm{max}} {\rm \bf e}^{Q} \right) ^2}} \right)^{q/2-1} \nonumber \\
&\cdot & \left[ 1 - \frac{1}{2} \cdot \frac{\left( 2 \frac{m_e c^2}{\hbar \omega} x_{\mathrm{max}} {\rm \bf e}^{Q} \right) ^2}{\sqrt{1 + \left( 2 \frac{m_e c^2}{\hbar \omega} x_{\mathrm{max}} {\rm \bf e}^{Q} \right) ^2} \cdot \left( 1 + \sqrt{1 + \left( 2 \frac{m_e c^2}{\hbar \omega} x_{\mathrm{max}} {\rm \bf e}^{Q} \right) ^2} \right)} \right] \\
&\cdot & \left[ \ln \left( \frac{0.684}{x_{\mathrm{max}}} \right) + \frac{1}{q} + \frac{1}{2(q-1)} \right], \ \ \ Q = \frac{2-3q}{2q(q-1)}. \nonumber
\end{eqnarray}

Following the same procedure as in previous section, we express the lowest value of momentum $p_{\mathrm{min}}$ in $x_{\mathrm{max}},$ in terms of ion kinetic energy $T_i$
\begin{equation}
x_{\mathrm{max}} = \hbar \omega \frac{A m_p}{m_e} \cdot \frac{E}{(c p_{\mathrm{min}})^2} = \hbar \omega \frac{A m_p}{m_e} \cdot \frac{A m_p c^2 + T_i}{T_i^2 + 2 A m_p c^2 T_i}
\end{equation}
and we define relation between kinetic energy of a particle and energy of an emitted photon as
\begin{equation}
T_i = \zeta_i \hbar \omega.
\end{equation}
Now $x_{\mathrm{max}}$ becomes
\begin{equation}
x_{\mathrm{max}} = \hbar \omega \frac{A m_p}{m_e} \cdot \frac{\zeta_i \hbar \omega + A m_p c^2}{(\zeta_i \hbar \omega)^2 + 2 \zeta_i \hbar \omega A m_p c^2} = \frac{A m_p}{m_e} \frac{1}{\zeta_i} \left( 1 + \frac{1}{1 + \frac{\zeta_i \hbar \omega}{A m_p c^2}} \right)^{-1}.
\end{equation}
As in the case of electrons, $\zeta_i$ is generally not constant and can be a function of $\hbar \omega.$ If $\zeta_i$ is defined, expression (44) with implemented (47) can be used as an analytical solution to ion non-thermal bremsstrahlung emission.

Somewhat simpler relations can be used when $\frac{\zeta_i \hbar \omega}{A m_p c^2} \ll 1$ and $x_{\mathrm{max}} \approx \frac{1}{2} \cdot \frac{A m_p}{m_e} \frac{1}{\zeta_i},$
that is, in the case of low photon energies. Relation (44) then reduces to
\begin{eqnarray}
\frac{\mathrm{d}W}{\mathrm{d}t \mathrm{d}\omega \mathrm{d}V} &=& \frac{16 Z^2 e^6}{3 m_e^2 c^4} n_e K_i (A m_p c^2)^{1-q/2} \left( 2 \zeta_i \hbar \omega \right)^{-q/2} \\
&\cdot & \frac{{\rm \bf e}^{Q(q/2-1)}}{2} \left[ \ln \left( 0.684 \cdot 2 \zeta_i \frac{m_e}{A m_p} \right) + \frac{1}{q} + \frac{1}{2(q-1)} \right]. \nonumber
\end{eqnarray}
When considering photons of high energies, where $\frac{\zeta_i \hbar \omega}{A m_p c^2} \gg 1$ and $x_{\mathrm{max}} \approx \frac{A m_p}{m_e} \frac{1}{\zeta_i},$ (44) can be approximated with
\begin{eqnarray}
\frac{\mathrm{d}W}{\mathrm{d}t \mathrm{d}\omega \mathrm{d}V} &=& \frac{16 Z^2 e^6}{3 m_e^2 c^4} n_e K_i \left( \zeta_i \hbar \omega \right)^{1-q} \\
&\cdot & {\rm \bf e}^{2Q(q/2-1)} \left[ \ln \left( 0.684 \cdot \zeta_p \frac{m_e}{A m_p} \right) + \frac{1}{q} + \frac{1}{2(q-1)} \right]. \nonumber
\end{eqnarray}
We see again, that there are two power laws and break point in the spectrum is determined from boundary between them as
\begin{equation}
\hbar \omega _b = \frac{A m_p c^2}{\zeta_i}.
\end{equation}

Parameters $\zeta_i$ and $\zeta_e,$ relation between them, their limits and spectral dependence are discussed in the theoretical framework in next section.

\section{Discussion and Conclusions}

We show that emissivity of electrons, as well as ions, beside spectral dependance, also depends on $\zeta_e$ and $\zeta_i,$ respectively. As said before, these coefficients can be functions of $\hbar \omega.$ We now derive relation between $\zeta_e$ and $\zeta_i,$ which clearly determines amount of contribution and shape of the curve of the individual scattering processes in overall emission.

We first discuss the scope of validity of Thomson cross section that we used in our derivation. Energy change of a photon scattering on electron in an arbitrary frame we derive from the law of conservation of the four-momentum $p^\mu_0 + k^\mu_0 = p^\mu_{\rm s} + k^\mu_{\rm s}:$
\begin{equation}
\omega _{\rm s} = \omega _0 \cdot \frac{\gamma (1 - \beta \cos \alpha)}{\gamma (1 - \beta \cos (\alpha + \theta)) + \frac{\hbar \omega}{m_e c^2} (1 - \beta \cos \theta)},
\end{equation}
where four-momentum vectors of electron and photon before scattering are represented by $p^\mu_0$ and $k^\mu_0,$ respectively and as $p^\mu_{\rm s}$ and $k^\mu_{\rm s}$ after the scattering process; $\beta = v/c$ and $v$ is the speed of electron. For simplicity, we assume that photon and electron are moving in the same plane, so that $\alpha$ is the angle between directions of photon and electron before and $\theta$ after the collision.

We see that, as long as $\hbar \omega \ll m_e c^2,$ the Compton effect is negligible. When photon energy becomes comparable with thermal energy of electron in plasma $\hbar \omega \sim k T_e ,$ then we must consider Inverse Compton scattering (ICs). It is most evident in the case of a head-on collision ($\alpha = \pi$) and deflection of a photon in the direction of an electron ($\theta = \pi$). Relation (51) then reduces to
\begin{equation}
\omega _{\rm s} = \frac{1 + \beta}{1 - \beta} \cdot \omega _0 .
\end{equation}
Concerning a lower limit of particle kinetic energy, $\beta$ is always close to zero for target electrons hit by virtual photons. Hence, ICs would become significant only in the case of very fast shocks ($v \sim 10000 \rm\ km/s$) with photon energy gain of $\sim 10 \ \%$ or in suprathermal plasma ($T_e \sim 10^6 \rm\ K$) when photon energy changes only by a few percents.

Although supernova remnants are strong synchrotron emitters, it is shown that in some cases \cite{r11} thermal bremsstrahlung may constitute an important part of the emission. The same may be true in the case of non-thermal bremsstrahlung, but at the end it is probably an indistinguishable mixture of both. Because of this, the most favorable environment for which detection of non-thermal bremsstrahlung would be possible is interstellar medium (ISM). With plasma temperatures $T_e \sim 10^4 \rm\ K$ and very low densities $n \sim 1 \rm\ cm^{-3},$ ISM provides conditions in which power law could be seen in photon spectra of cosmic-ray particles down to photon energies of $\sim 0.1 \rm\ eV.$ Below this energy, the spectra would be composite of a thermal plasma emission and non-thermal cosmic-ray radiation. Magnetic field, as a constituent of ISM, can capture cosmic rays of lower momenta without modifying their power law distribution and hence, improve non-thermal bremsstrahlung emission, if magnetic diffusion is slow enough.

Therefore, there are two limits in our description of the photon spectra. The first one is physical -- the thermal energy of target plasma from the low photon energy side of the spectrum, and the second is a constraint of the theory -- electron rest energy or Thomson's limit, from the upper side of the spectrum. Nevertheless, inside this range, scattering can be fully considered as Thomson.

We now derive relations for $\zeta_e$ and $\zeta_i$ from the two cases of bremsstrahlung emission and we also impose some constraints. As previously, we define variables as primed, meaning that electron is in rest in that frame. In the case of relativistic electron bremsstrahlung, kinetic energies of electron and ion are defined as
\begin{eqnarray}
T_e &=& \zeta_e \hbar \omega = (\gamma -1) m_e c^2 \Rightarrow \gamma = 1 +\zeta_e \frac{\hbar \omega}{m_e c^2}, \nonumber \\
T_i ' &=& \zeta_i \hbar \omega ' = (\gamma -1) A m_p c^2 \Rightarrow \zeta_i = \frac{A m_p}{m_e} \zeta_e \frac{\omega}{\omega '} = \frac{A m_p}{m_e} \zeta_e \gamma. \nonumber
\end{eqnarray}
Similarly, in the case of relativistic ion bremsstrahlung we derive kinetic energies as
\begin{eqnarray}
T_i ' &=& \zeta_i \hbar \omega ' = (\gamma -1) A m_p c^2 \Rightarrow \gamma = 1 +\zeta_i \frac{\hbar \omega '}{A m_p c^2}, \nonumber \\
T_e &=& \zeta_e \hbar \omega = (\gamma -1) m_e c^2 \Rightarrow \zeta_e = \frac{m_e}{A m_p} \zeta_i \frac{\omega '}{\omega} = \frac{m_e}{A m_p} \zeta_i \frac{1}{\gamma}. \nonumber
\end{eqnarray}
In both cases, relation between $\zeta_e$ and $\zeta_i$ is derived as
\begin{equation}
\zeta_i = \frac{A m_p}{m_e} \zeta_e \gamma
\end{equation}
and we define $\gamma$ in two ways, depending on whether the emission comes from ion or electron rest frame
\begin{eqnarray}
\gamma &=& 1 + \zeta_e \frac{\hbar \omega}{m_e c^2}, \\
\gamma &=& 1 + \zeta_i \frac{\hbar \omega '}{A m_p c^2}.
\end{eqnarray}
If $\zeta_e$ is defined, we derive $\zeta_i$ from (53) with implementation of (54)
\begin{equation}
\zeta_i = \frac{A m_p}{m_e} \zeta_e \left( 1 + \zeta_e \frac{\hbar \omega '}{m_e c^2} \gamma \right) = \frac{A m_p}{m_e} \zeta_e \left[ 1 + \zeta_e \frac{\hbar \omega '}{m_e c^2} + \left( \zeta_e \frac{\hbar \omega '}{m_e c^2} \right)^2 +\ldots \right], \nonumber
\end{equation}
which converges only if $\zeta_e \frac{\hbar \omega '}{m_e c^2} < 1$ and in this case we get a geometric progression
\begin{equation}
\zeta_i = \frac{A m_p}{m_e} \cdot \frac{\zeta_e}{1 - \zeta_e \frac{\hbar \omega '}{m_e c^2}}.
\end{equation}
If $\zeta_i$ is defined, similar to previously, we derive $\zeta_e$ from (53) again by implementing (54), because this time we want to express it as function of photon energy in a rest frame of an ion
\begin{eqnarray}
\zeta_i &=& \frac{A m_p}{m_e} \zeta_e \cdot \left( 1 + \zeta_e \frac{\hbar \omega}{m_e c^2} \right), \nonumber \\
\zeta_e &=& \frac{m_e c^2}{\hbar \omega} \cdot \frac{\sqrt{1 + 4 \zeta_i \frac{\hbar \omega}{A m_p c^2}} - 1}{2},
\end{eqnarray}
which is now convergent for any photon energy. This implies that it is more convenient to define $\zeta_i$ as a constant and then to express $\zeta_e$ through (57). In a sense of physical approach, this resembles the process of scattering of virtual photon field carried by an ion, on electron in rest. Thomson cross section is also defined in a similar manner. We now rewrite the set of equations for particles kinetic energy as
\begin{eqnarray}
T_i &=& \zeta_i \hbar \omega_{\mathrm{lab}},\ \ \ (\omega_{\mathrm{lab}} = \omega '), \\
T_e &=& m_e c^2 \cdot \frac{\sqrt{1 + 4 \zeta_i \frac{\hbar \omega_{\mathrm{lab}}}{A m_p c^2}} - 1}{2},\ \ \ (\omega_{\mathrm{lab}} = \omega),
\end{eqnarray}
because we need to express kinetic energy of a particle as a function of photon energy ($\hbar \omega_{\mathrm{lab}}$) in laboratory frame which, to a high degree of accuracy, is the same as the observer frame.

If we now define kinetic energies using these relations and dependence (57), we see that in the case of relativistic ion bremsstralung, relations (44), (47), (48), (49) and (50) remain unchanged because $\zeta_i$ is constant. However, concerning relations of relativistic electron bremsstrahlung, $\zeta_e$ now has weak nonlinear dependence on photon energy and we have to include it in (25), (26), (27), (28) and (29). The most important result is that the amount of contribution of these scattering processes in overall emission is now determined. For most photon energies, relation $\zeta_i \approx \frac{A m_p}{m_e} \zeta_e$ applies, so that ratio of electron to ion bremsstrahlung emissivity can be roughly estimated as
\begin{equation}
\frac{\varepsilon ^e _\nu}{\varepsilon ^i _\nu} \approx \left( \frac{A m_p}{m_e} \right)^{(q-1)/2},
\end{equation}
but at high energies, nonlinear dependence additionally flattens
the power law. The most important feature in photon spectra is
that, in low energy part, curves of these two types of non-thermal
bremsstrahlung are nearly parallel with mutual dependance
$\varepsilon ^{e/i} _\nu \sim (\hbar \omega)^{-q/2}$ and spectral
index $\alpha = q/2.$ After the break points defined in (29) and
(50), power laws of electron and ion bremsstrahlung in high energy
part of photon spectra are given by $\varepsilon ^e _\nu \sim
(\hbar \omega)^{(1-q)/2}$ and $\varepsilon ^i _\nu \sim (\hbar
\omega)^{1-q},$ respectively. Logarithmic dependance is also
present in both cases, as seen in (28) and (49), but for photon
energies lower than $511 \rm\ keV$ it is insignificant.

Relativistic protons and heavier ions, as well as electrons, the main constituents of cosmic rays that fill out our Galaxy and that we observe on Earth, are believed to be accelerated at strong shocks of supernova remnants and in other high-energy astrophysical processes. While all particle share a similar energy index $q\sim 2.5-2.7$, the cosmic rays are dominated by protons. Observed ratio $K_p/K_e$ is about 100. This is in strong contrast to cosmic abundances, H:He = 10:1, implying $n_p \approx n_e \equiv n$ for full ionization. Hence, we use these facts to estimate emission coefficients for relativistic electron and relativistic proton bremsstrahlung. Full relations (7) and (32) we solve numerically and plot them together with derived analytical expressions (26) and (44) in Figure 1.

\begin{figure}[!h]
\centerline{\psfig{file=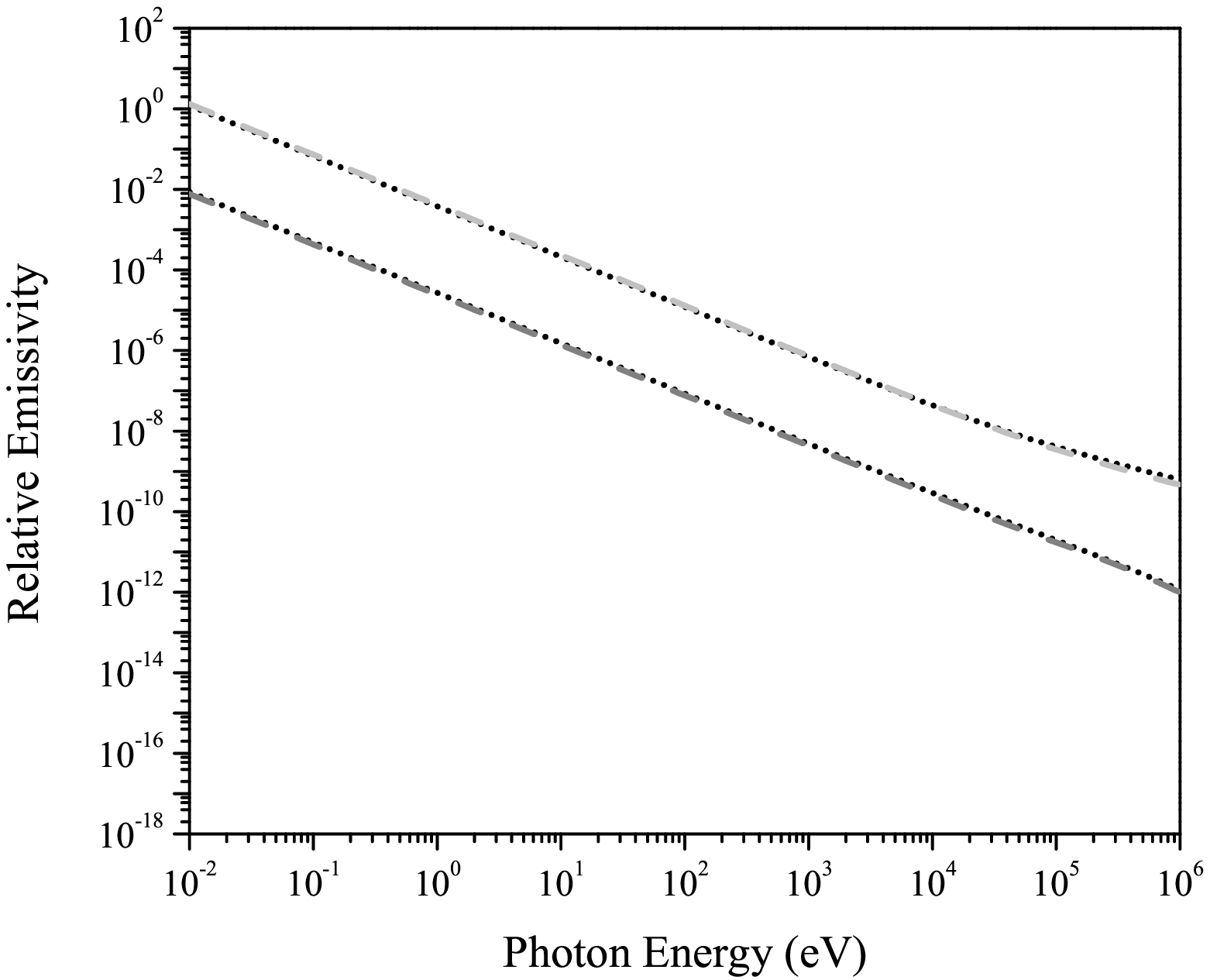,width=0.49\textwidth} \psfig{file=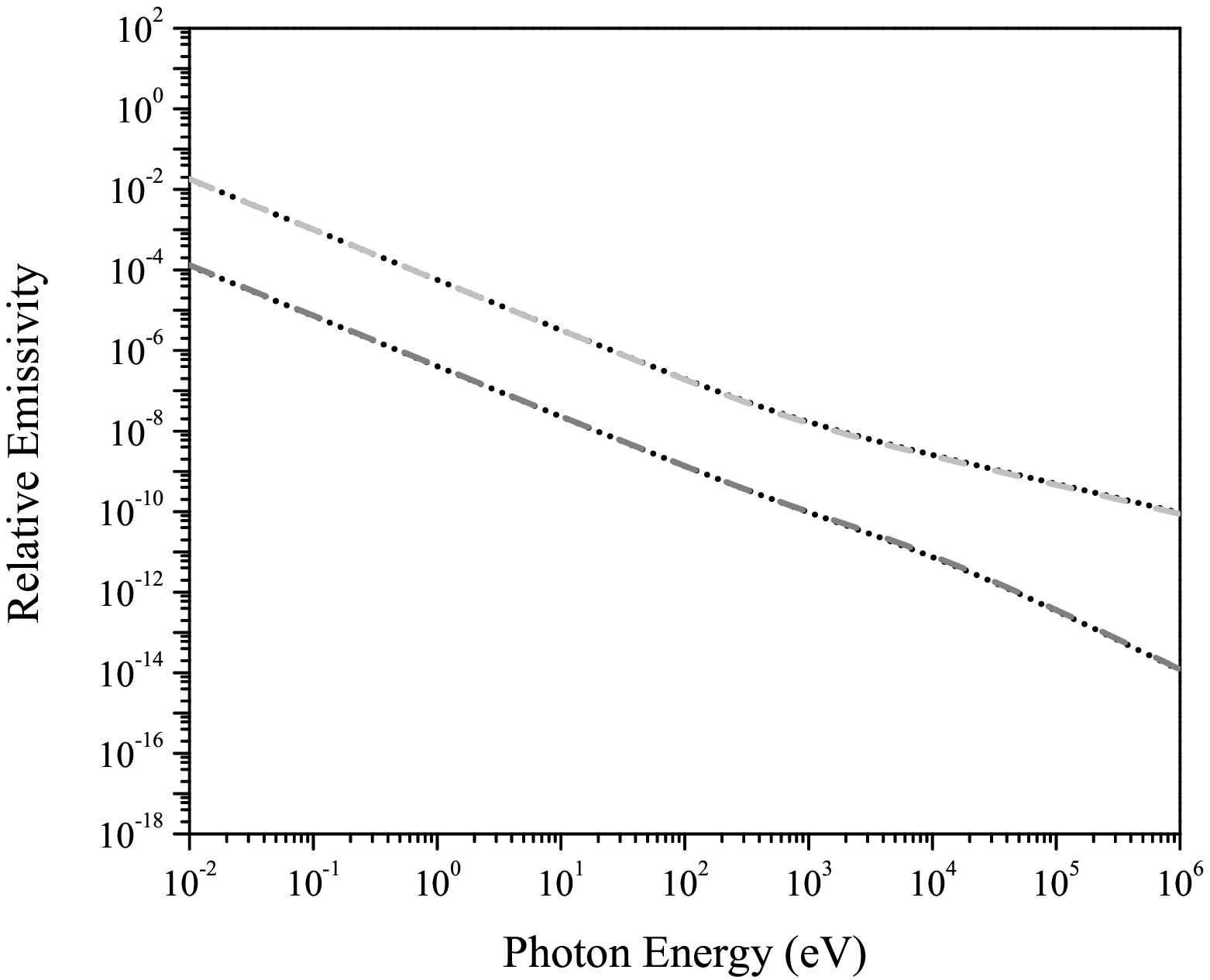,width=0.49\textwidth}}
\centerline{\psfig{file=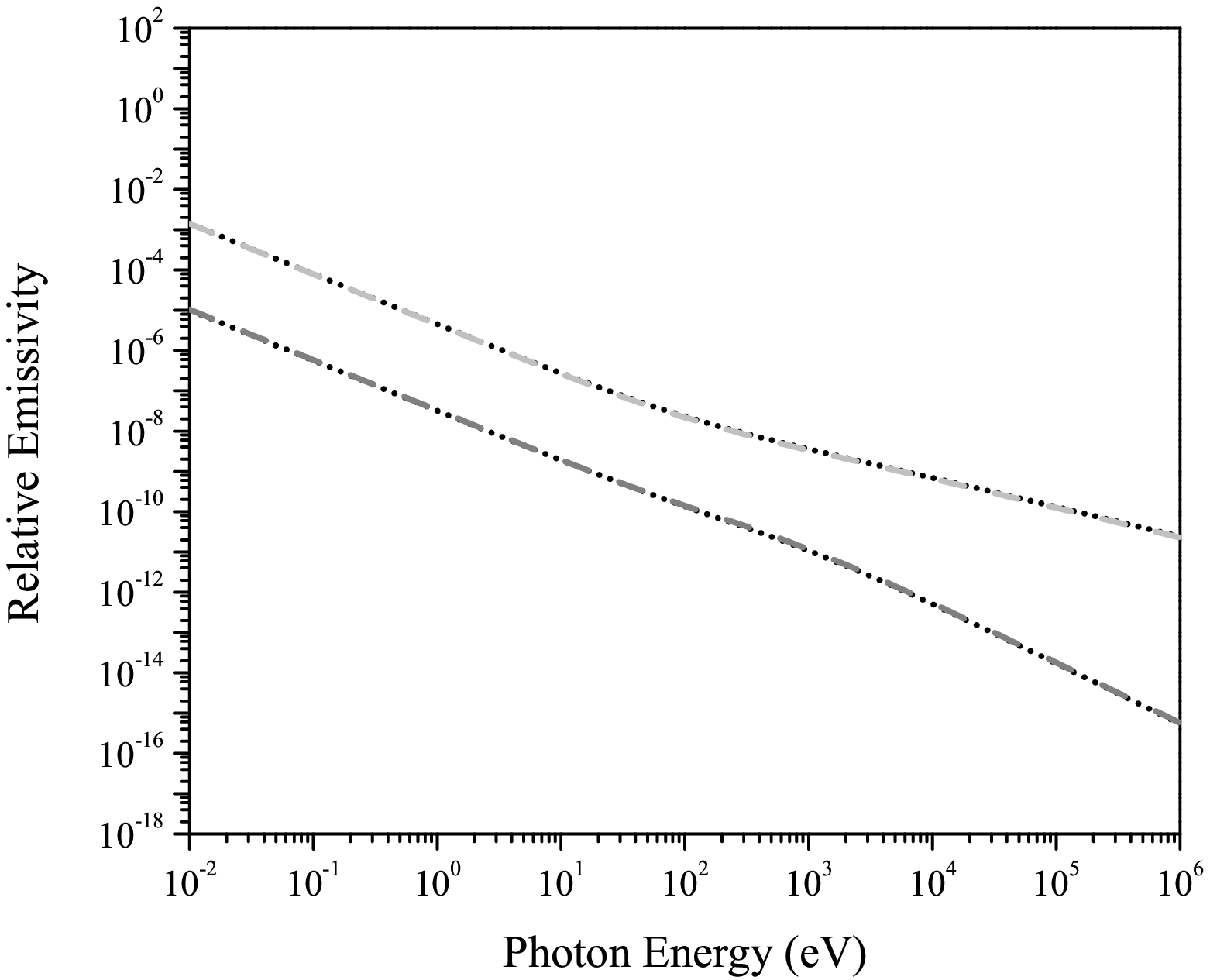,width=0.49\textwidth} \psfig{file=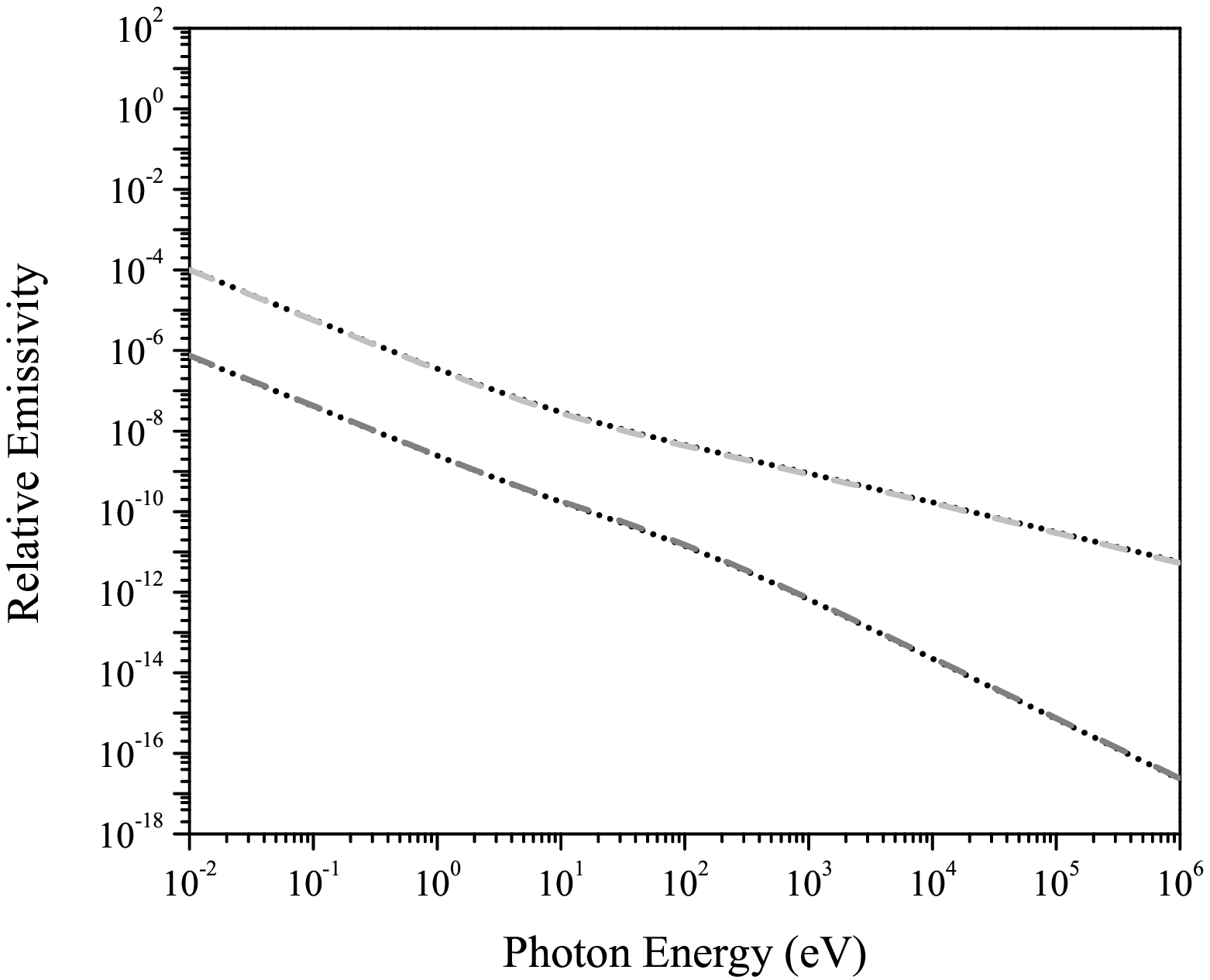,width=0.49\textwidth}}
\vspace*{8pt}
\caption{Emissivity of relativistic electron and proton (inverse) bremsstrahlung in relative units ($\varepsilon ^{i/e}_\nu = \frac{8 Z^2 e^6}{3 m_e^2 c^4} n K_e \times \mathrm{Relative{\ }Emissivity}$) vs. energy of emitted photons in $\rm eV$, considered for momentum distribution of particles with energy index $q=2.5$ and $\zeta_p=1$ (upper left), $\zeta_p=10^2$ (upper right), $\zeta_p=10^3$ (lower left) and $\zeta_p=10^4$ (lower right). Exact numerical solutions, with $x _{\mathrm{min}}$ set to zero, are plotted as black dotted lines and analytical approximations as light gray and dark gray dashed lines, for electrons and protons, respectively. Ratio $(K_i/K_e) \approx (A m_p/m_e)^{(q-1)/2}$ is used and is included in relative emissivity of proton bremsstrahlung.}
\end{figure}

While electron-ion bremsstrahlung can generate photons at all energies up to $T_{e/i},$ the maximum energy of a photon emitted in electron-electron scattering ($\epsilon _{\mathrm{max}} =\hbar \omega
_{\mathrm{max}}$), tends to zero for ultra-relativistic cosmic-ray
electrons, except when $\theta = 0.$ For highly non-relativistic
electrons, $\epsilon _{\mathrm{max}} \rightarrow T_e / 2$ for all
scattering angles. In the case of mildly relativistic cosmic-ray
electrons, $\epsilon _{\mathrm{max}}$ is always less than $T_e$
\cite{r16}. Despite these facts, we see that the most weighted
part of momentum distribution is the low energy part, for which
the value of $\epsilon _{\mathrm{max}}$ can constitute a
significant part of $T_e.$ This could indicate that
electron-electron bremsstrahlung might also be of importance, at
least to some extent or even more important than relativistic ion
bremsstrahlung.

Also, synchrotron emission, which is typical for radio
domain, in some sources such as young supernova remnants, can
reach X-ray part of the spectrum. In the X-rays instead of
emissivity one usually considers photon number distribution
$n(\hbar\omega)\propto (\hbar\omega)^{\mathnormal{-\Gamma}}$ where
photon index is $\mathnormal{\Gamma} =\alpha +1$ \cite{r9}.
Typically, X-ray synchrotron spectra of supernova remnants have
rather steep indices, $\mathnormal{\Gamma} = 2-3.5$ \cite{r10}.
Since in the case of synchrotron emission $\alpha = \frac{q-1}{2}$
\cite{r6,r7}, this can only be explained by significant steepening
of the electron energy distribution, presumably due to the
radiative losses. Nevertheless, non-thermal bremsstrahlung of
mildly relativistic electrons with $\alpha =q/2$ can not be
excluded for lower observed $\mathnormal{\Gamma}$.

As the subject of further work, we will include a full Compton cross section in equations, in order to extend the theory to all photon energies, as well as to recalculate cross section for electron-electron bremsstrahlung and also to review the method of virtual quanta in light of quantum fields.

\section*{Acknowledgments}

During the work on this paper the authors were financially supported by the\\ Ministry of Education, Science and Technological Development of the Republic of Serbia through the projects: 176005 "Emission nebulae: structure and evolution" (VZ, BA, AD, MP) and 176004 "Stellar physics" (BA).

%


\end{document}